\documentstyle[12pt,,dina4,epsfig]{article}

\newcommand{\stern}{+\hspace*{-0.327cm}\times}
\def\kreiso{\lower1.5pt\hbox{\Large $\bullet$}}
\def\kreisv{\raise1.5pt\hbox{$\scriptstyle\bigcirc$}}
\begin{document}

\vspace{-3.5cm}
\begin{flushleft}
{\normalsize DESY 97-217} \hfill\\
{\normalsize HUB-EP-97/85} \hfill\\
{\normalsize November 1997} 
\end{flushleft}
\vspace{0.5cm}

\begin{center}
{\LARGE\bf Spin Structure Functions from \\[0.3em] 
  Lattice QCD$\footnote{Talk given by G. Schierholz
     at the
    $2^{\scriptsize\mbox{nd}}$
  International Workshop
  {\it Deep-Inelastic Scattering off Polarized Targets: Theory Meets 
  Experiment}, Zeuthen, September 1-5, 1997}$}

\vspace{1cm}
{M. G\"ockeler$^a$, R. Horsley$^b$, H. Perlt$^c$, P. Rakow$^d$,\\ 
G. Schierholz$^{d,e}$, A. Schiller$^c$ and P. Stephenson$^d$}

\vspace*{1cm}
{\it $^a$Institut f\"ur Theoretische Physik, Universit\"at Regensburg,
D-93040 Regensburg}\\

\vspace*{3mm}
{\it $^b$Institut f\"ur Physik, Humboldt-Universit\"at zu Berlin,
D-10115 Berlin}\\

\vspace*{3mm}
{\it $^c$Institut f\"ur Theoretische Physik, Universit\"at Leipzig,
D-04109 Leipzig}\\

\vspace*{3mm}
{\it $^d$Deutsches Elektronen-Synchrotron DESY, Institut f\"ur
  Hochenergiephysik and HLRZ, D-15735 Zeuthen}\\

\vspace*{3mm}
{\it $^e$Deutsches Elektronen-Synchrotron DESY, D-22603 Hamburg}\\

\vspace*{2cm}

\end{center}

\begin{abstract}
We report on new results of the spin dependent structure
functions $g_1$ and $h_1$ of the nucleon. An attempt is made to convert
the moments, which is what one computes on the lattice, to quark
distribution functions. 
\end{abstract}

\vspace*{1cm}

\section{Introduction}

In the past polarization data have often been the graveyard of fashionable
models. Measurements of the polarized deep-inelastic structure
functions of the nucleon over the last decade have once again borne out 
this experience.
While the naive quark parton model has been very successful in predicting
the gross features of hadrons, it was a surprise to find that
it fails to explain the spin properties of the nucleon.

There is significant interest in an {\it ab initio} calculation of the
nucleon structure functions now. The theoretical framework of such a
calculation 
is the operator product expansion. While the Wilson coefficients can
be computed perturbatively, the hard part of the
calculation is the determination of the forward nucleon matrix
elements of the operators. This is a non-perturbative problem, and the
technique to solve it is lattice QCD. For recent work on the subject
see~\cite{us}. A confrontation of the 
experimental data with the lattice 
predictions will be a crucial test of our understanding of the structure
of the nucleon. 

At the twist two level the quark sector of the nucleon is completely
specified by the spin averaged structure functions $F_1(x,Q^2)$, 
$F_2(x,Q^2)$ and the polarized structure functions $g_1(x,Q^2)$,
$h_1(x,Q^2)$. In this talk we will report on new results of the
polarized structure functions. 

The talk is organized as follows. In sec.~2 we introduce the notation.
In sec.~3 we give the results for the moments of $g_1$ and $h_1$. We
furthermore give an extrapolation of the axial vector coupling
of the nucleon, $g_A$, to the continuum limit. 
The moments can be converted into real structure functions by an inverse
Mellin transform. We will present first results on $g_1(x,Q^2)$ in
sec.~4. Finally, in sec.~5 we conclude.

\section{Basics}

The structure functions $g_1$ and $h_1$
have simple parton model interpretations. The structure function $g_1$
measures the quark helicity distribution in a longitudinally polarized
nucleon, while $h_1$ measures the probability of finding a quark in a spin
eigenstate of the operator $\not{\! s}_\bot \gamma_5$ in a transversely
polarized nucleon.  

In the following we shall consider only {\it non-singlet} structure
functions and distributions. Only these distribution
functions are accessible in quenched lattice QCD.

Let us first consider the structure function $g_1$. If the sea is
assumed to be flavor symmetric, the leading twist contribution can be written
\begin{equation}
g_1(x,Q^2) = \frac{1}{2} \sum_f \int_x^1 \frac{\mbox{d}y}{y}
c_1^{(f)}\left(\frac{x}{y},\frac{Q^2}{\mu^2}\right) \Delta q^{(f)}(y,\mu),
\label{g1}
\end{equation}
where 
\begin{equation}
\Delta q^{(f)}(x,\mu) = q^{(f)}_\uparrow(x,\mu) - q^{(f)}_\downarrow(x,\mu),
\end{equation}
and $q^{(f)}_{\uparrow (\downarrow)}$ is the probability distribution
of a quark 
of flavor $f$ and spin parallel (anti-parallel)
to the parent spin of the nucleon. The so-called splitting functions 
$c^{(f)}_1$ are determined by the Wilson coefficients
\begin{equation}
c^{(f)}_{1,n}\left(\frac{Q^2}{\mu^2}\right) =
\int_0^1 \mbox{d}x x^n c_1^{(f)}\left(x,\frac{Q^2}{\mu^2}\right) 
\end{equation}
through an inverse Mellin transform.
Similarly, the distribution functions can be derived from their moments, 
\begin{equation}
\Delta^{(n)} q^{(f)}(\mu) = 
\int_0^1 \mbox{d}x x^n \Delta q^{(f)}(x,\mu).
\label{Del}
\end{equation}
We denote the lowest moment by
\begin{equation}
\Delta q^{(f)} \equiv \Delta^{(0)} q^{(f)}.
\end{equation}
According to the operator product expansion the moments are given by forward
nucleon matrix elements of local
operators. For $\Delta^{(n)} q^{(f)}$ we have
\begin{equation}
\langle\vec{p},\vec{s}|{\cal O}_{\{\sigma \mu_1 \cdots
  \mu_n\}}^{5(f)}|\vec{p},\vec{s}\rangle = \frac{2}{n+1} \Delta^{(n)}
q^{(f)} [s_\sigma p_{\mu_1} \cdots p_{\mu_n} + \cdots - \mbox{traces}],
\label{axmat}
\end{equation}
where 
\begin{equation}
{\cal O}_{\{\sigma \mu_1 \cdots \mu_n\}}^{5(f)} =
\left(\frac{\mbox{i}}{2}\right)^n \bar{q}^{(f)}\gamma_\sigma \gamma_5
\stackrel{\leftrightarrow}{D}_{\mu_1} \cdots
\stackrel{\leftrightarrow}{D}_{\mu_n} q^{(f)} - \mbox{traces}.
\label{ax}
\end{equation}
Here $\{\cdots\}$ means symmetrization.
The lowest moment $\Delta q^{(f)}$ measures the {\it axial vector charge}
of the nucleon.

Similar expressions can be derived for the structure function 
$h_1$~\cite{jaffe}. One
simply has to replace  $\Delta q^{(f)}(x,\mu)$ in eq.~(\ref{g1}) by
\begin{equation}
\delta q^{(f)}(x,\mu) = q^{(f)}_\bot(x,\mu) - q^{(f)}_\top(x,\mu),
\end{equation}
where $q^{(f)}_{\bot(\top)}$ is the probability distribution of a
quark of flavor 
$f$ and spin $\not{\! s}_\bot \gamma_5$ parallel
(anti-parallel) to the spin of the nucleon, giving
\begin{equation}
\delta^{(n)} q^{(f)}(\mu) = 
\int_0^1 \mbox{d}x x^n \delta q^{(f)}(x,\mu).
\label{del}
\end{equation}
We denote the lowest moment by
\begin{equation}
\delta q^{(f)} \equiv \delta^{(0)} q^{(f)}.
\end{equation}
The moments are given by the matrix elements
\begin{equation}
\langle\vec{p},\vec{s}|{\cal O}_{\sigma\{\mu_1 \cdots
  \mu_n\}}^{(f)}|\vec{p},\vec{s}\rangle = \frac{2}{m_N} \delta^{(n-1)}
q^{(f)} [(s_\sigma p_{\mu_1} - s_{\mu_1} p_\sigma) p_2 \cdots p_{\mu_n} 
+ \cdots - \mbox{traces}],
\end{equation}
where 
\begin{equation}
{\cal O}_{\sigma \{\mu_1 \cdots \mu_n\}}^{(f)} =
\left(\frac{\mbox{i}}{2}\right)^{n-1}
\bar{q}^{(f)}\sigma_{\sigma\mu_1}
\gamma_5 \stackrel{\leftrightarrow}{D}_{\mu_2} \cdots
\stackrel{\leftrightarrow}{D}_{\mu_n} q^{(f)} - \mbox{traces}.
\label{ten}
\end{equation}

The lattice operators, which are computed at the scale $1/a$ ($a$:
lattice constant), must be renormalized and brought into the same scheme
in which the Wilson coefficients have been calculated. Generically we
can write
\begin{equation} 
{\cal O}_i(\mu) = Z_{i j} {\cal O}_j(a),
\end{equation}

\clearpage
\begin{figure}[h]
\begin{centering}
\epsfig{figure=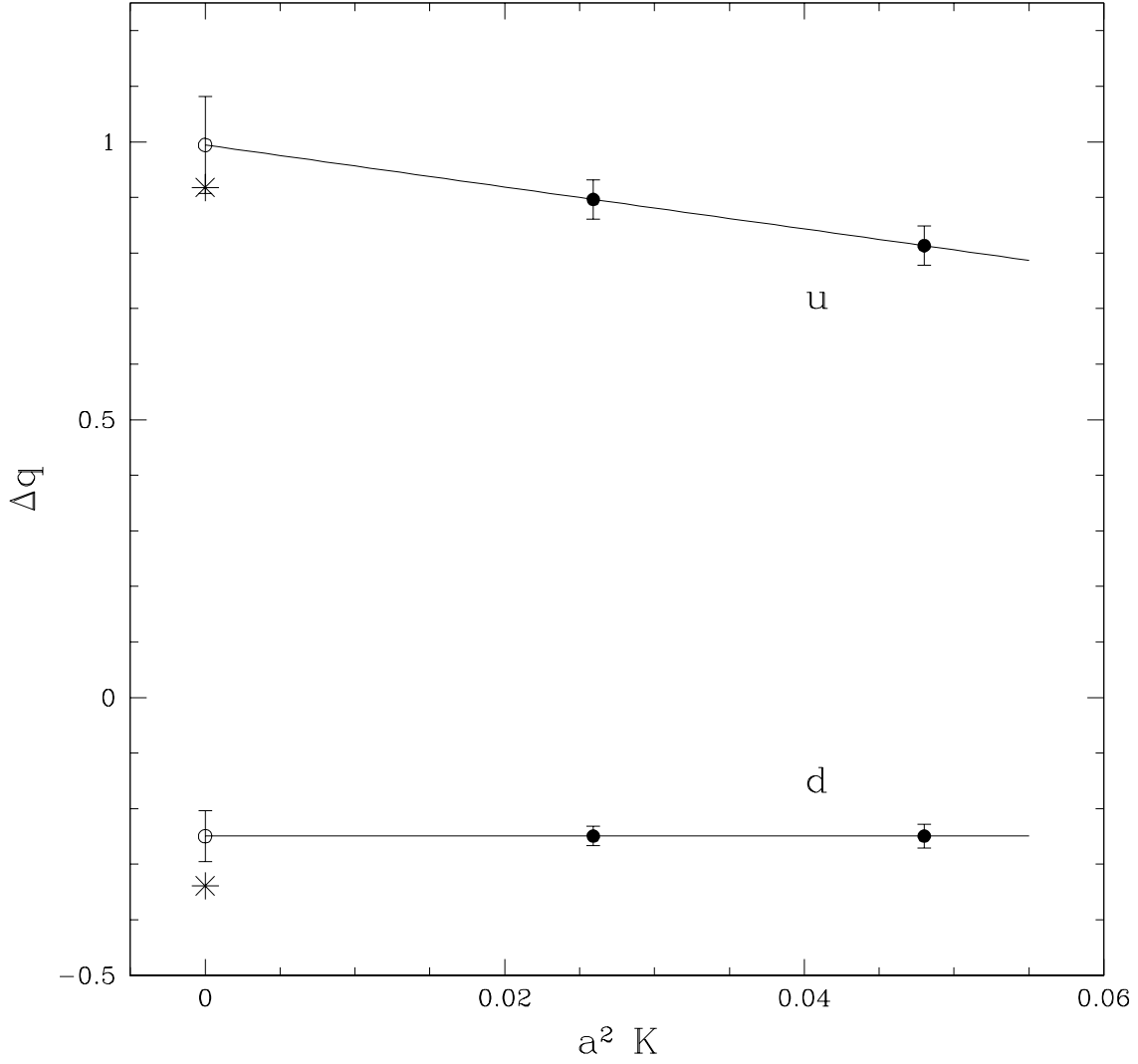,height=15.5cm,width=15.5cm}
\caption{The quenched moments $\Delta u$ and $\Delta d$ plotted as a 
function of
$a^2$. The lattice spacing is given in units of the string tension,
$K$. The lattice data are denoted by $\kreiso$, the extrapolated
values by $\kreisv$.  
The phenomenological 
values~\protect\cite{gs} are denoted by $\stern$.}
\label{deltaud}
\end{centering}
\end{figure}

\clearpage
\noindent
where the indices distinguish between the various lattice operators 
which can (and do) mix under renormalization. It is a task of its own to
compute the renormalization constants accurately. For details of the
calculation we refer
the reader to the literature~\cite{ren} and the talk of 
Schiller~\cite{arwed}. 

The lowest moment $\delta q^{(f)}$ measures the {\it tensor charge} of 
the nucleon. In the non-relativistic quark model axial vector and
tensor charges are equal, i.e. $\delta q^{(f)} = \Delta q^{(f)}$. 
The structure function $h_1$ has opposite chiral properties to $g_1$. It
can be measured in polarized Drell-Yan processes, but not in
deep-inelastic lepton-hadron scattering. For other ideas see~\cite{rlj}.

\section{Moments}

The quantities to be computed on the lattice are the moments
(\ref{Del}) and (\ref{del}). We have new results on $\Delta q^{(f)}$ and
$g_A$, on $\Delta^{(1)} q^{(f)}$ and on $\delta q^{(f)}$. 
In the quenched approximation, which we are using, $f = u, d$, and we
write $q^{(u)} = u$ and $q^{(d)} = d$. In the following we drop the
argument $\mu$ from the quark distributions.

\subsection*{$\Delta q$}

In lattice calculations one imagines the high frequency modes higher
than the cut-off being integrated out. The logarithmically singular
contributions of these modes are absorbed into the bare parameters,
such as the coupling constant, while the power-behaved contributions 
are usually left unaccounted for. For Wilson fermions,
which are widely used, the power corrections are of $O(a)$. By improving
the action, these corrections can be reduced to $O(a^2)$~\cite{imp}, 
thus giving results which are closer to the continuum limit $a = 0$. 

Removing $O(a)$ effects from the matrix elements requires improving
the operators as well. For the axial vector current the improved
operator, and its renormalization constant, have been derived
recently~\cite{alpha}. We will make use of these results here.

We have done calculations at two values of the coupling, $\beta = 6.0$
and $6.2$, corresponding to lattice spacings of $a \approx 0.1$ and
$0.07 \;\mbox{fm}$, respectively. The results for $\Delta u$ and 
$\Delta d$ are plotted in Fig.~\ref{deltaud}. The values given refer to
the chiral limit, i.e. zero quark mass, which are obtained from the
lattice data by a suitable extrapolation.

As the remaining errors
are of $O(a^2)$, we may fit the cut-off dependence by a formula of
the form $c_0 + c_2 a^2$, and use this formula to extrapolate the
result to the continuum ($a = 0$) limit. The outcome is shown by the
solid lines. Clearly, we would have liked to have at least one more data point
at another value of the coupling. We are working on that. We compare
our results with the phenomenological valence quark distribution
functions~\cite{gs}. The agreement is good, considering that the
errors on the phenomenological values are of the order of $10 \%$.

\clearpage
\begin{figure}[h]
\begin{centering}
\epsfig{figure=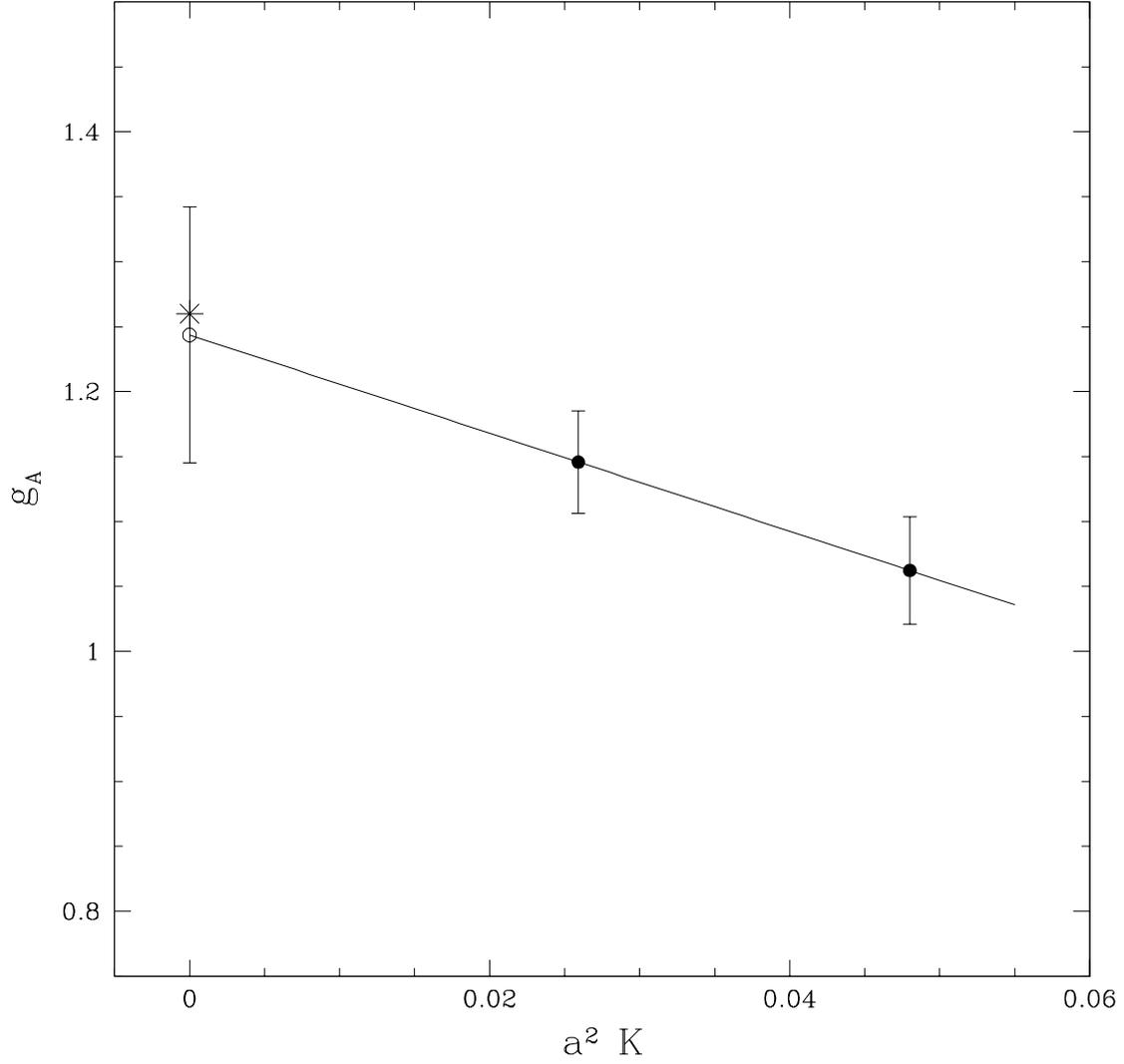,height=15.5cm,width=15.5cm}
\caption{The axial vector coupling of the nucleon $g_A$ as a function of
$a^2$. The lattice spacing is given in units of the string tension,
$K$. The lattice results are denoted by $\kreiso$, the extrapolated
values by $\kreisv$.  
The experimental value is denoted by $\stern$.}
\label{ga}
\end{centering}
\end{figure}

\clearpage

\subsection*{$g_A$}

A quantity, which is known very precisely experimentally, is the axial
vector coupling of the nucleon,
\begin{equation}
g_A = \Delta u - \Delta d = 1.26.
\end{equation}
In Fig.~\ref{ga} we show our results for $g_A$, together with the
extrapolation of the lattice data
to the continuum limit. We find the continuum result to be in good
agreement with the experimental value. 

This figure indicates quite clearly how important it is to correct for
finite cut-off effects. Even after improving the action and the
operators, cut-off corrections can be
quite substantial still. We see
that the lattice result increases by approximately $20 \%$ going from our
coarsest lattice at $\beta = 6.0$ to the continuum limit. 

\subsection*{$\delta q$}

\begin{figure}[h]
\begin{centering}
\epsfig{figure=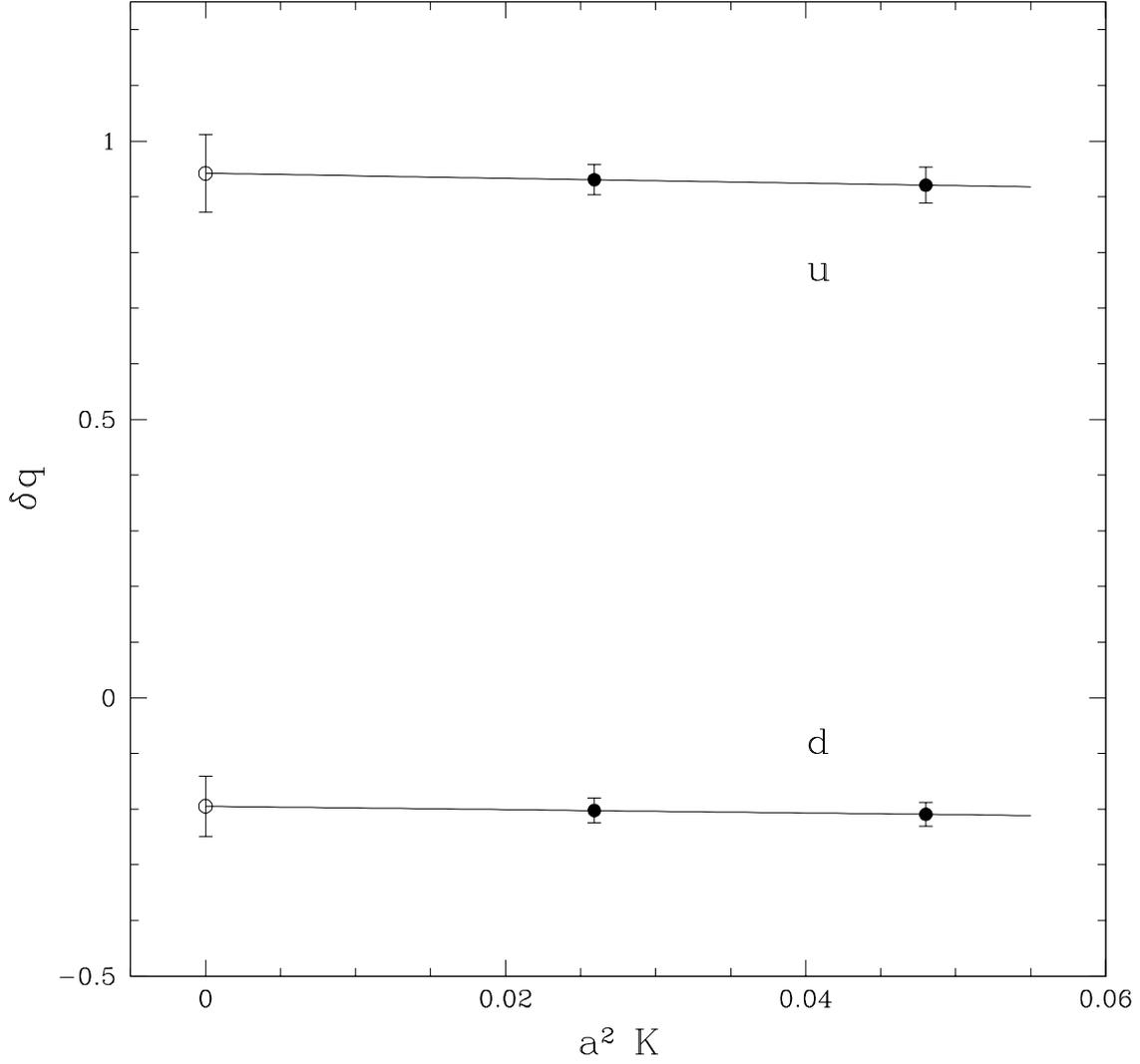,height=15.5cm,width=15.5cm}
\caption{The quenched moments $\delta u$, $\delta d$ as a function of 
$a^2$. The lattice spacing is given in units
of the string tension, $K$. The lattice results are denoted by $\kreiso$,
the extrapolated values by $\kreisv$. The numbers are
renormalized at the scale $\mu^2 = 4 \; \mbox{GeV}^2$.}
\label{ddel}
\end{centering}
\end{figure}

The moments $\delta u$, $\delta d$ have been computed using the
improved action, but so far the operator has not been improved. We are
currently working on this problem. For
the renormalization constant we have taken the tadpole
improved~\cite{lm} one-loop perturbative
result~\cite{lat96}. This means that the $O(a)$ corrections have not 
completely been removed in this case, as opposed to the previous case. 
The result of the calculation is shown in Fig.~\ref{ddel}. This work
adds another value of $\beta$ to our previous
calculation~\cite{ppaul}.

Before we discuss the result,
let us find out what the tensor charge actually tells us
about the structure of the nucleon. For a stationary nucleon the
operator (\ref{ten}) differs from (\ref{ax}) by a factor of $\gamma_0$.
In the non-relativistic limit fermions are in eigenstates of $\gamma_0$,
and so $\delta q$ and $\Delta q$ are equal. By comparing $\delta q$ and 
$\Delta q$ for a real nucleon, we can gain insight into how relativistic
the constituents are. 

Comparing $\delta q$ and $\Delta q$ in Figs.~\ref{deltaud}, \ref{ddel} 
now, we see that they are equal within
the error bars. This shows that a non-relativistic description of the
spin structure of the nucleon is quite adequate. Does this mean that the
quarks are in a relative s-wave and the missing spin is coming from the
gluons and the sea quarks? Further investigations will have to show.

Because the operator (\ref{ten}) with $n = 1$ is odd under charge
conjugation, sea quarks do not contribute to $\delta q$. This means that
we might hope that the quenched calculation is
giving an answer close to the true value.

\subsection*{$\Delta^{(1)} q$}

Another quantity which receives contributions from the valence quarks 
only is the second moment $\Delta^{(1)} q$. This moment is found from 
the $n = 1$ case of eqs.~(\ref{axmat}), (\ref{ax}). Again, the reason 
is that the operator is odd under charge conjugation. 

\clearpage

\clearpage

The special feature of this moment is that it is directly accessible
experimentally, which allows us to test the valence quark distribution
without further phenomenological analysis.

For the polarization asymmetry~\cite{lech} of $\pi^+$ minus $\pi^-$
inclusive cross sections one finds to lowest order in $\alpha_s$
\begin{equation}
A_p^{\pi^+ - \pi^-} = \frac{4 \Delta u^{\scriptsize\mbox{val}}(x) - 
\Delta d^{\scriptsize\mbox{val}}(x)}{4 u^{\scriptsize\mbox{val}}(x) - 
d^{\scriptsize\mbox{val}}(x)}
\label{ap}
\end{equation}
for a proton target, and 
\begin{equation}
A_d^{\pi^+ - \pi^-} = \frac{\Delta u^{\scriptsize\mbox{val}}(x) + 
\Delta d^{\scriptsize\mbox{val}}(x)}{u^{\scriptsize\mbox{val}}(x) + 
d^{\scriptsize\mbox{val}}(x)}
\label{ad}
\end{equation}
for a deuteron target. The fragmentation functions, as well as the sea
quark contributions, drop out because of isospin invariance relating the
various fragmentation functions with each other. 

The polarization asymmetries have been measured by the
SMC-Collaboration. For the lowest moment of the valence quark
distribution they found~\cite{smc} 
$\Delta u^{\scriptsize\mbox{val}} = 1.01 \pm 0.19 \pm 0.14$ and
$\Delta d^{\scriptsize\mbox{val}} = - 0.57 \pm 0.22 \pm 0.11$. 
The experimental 
errors are still a little too large to make a quantitative comparison
with the lattice results. The SMC-Collaboration has
recently extended their analysis to the second moment~\cite{pretz}. 
At $\mu^2 = 10 \: \mbox{GeV}^2$ they obtain the result
\begin{eqnarray}
\Delta^{(1)} u^{\scriptsize\mbox{val}} &=&  0.169 \pm 0.018 \pm 0.012, \\
\Delta^{(1)} d^{\scriptsize\mbox{val}} &=& -0.055 \pm 0.027 \pm 0.011.
\end{eqnarray}
A recent lattice calculation, renormalized at the same scale, 
gives~\cite{lech2}
\begin{eqnarray}
\Delta^{(1)} u &=&  0.189 \pm 0.08, \\
\Delta^{(1)} d &=& -0.0455 \pm 0.0032.
\end{eqnarray}
The lattice and experimental results agree within their respective
errors.  

\section{Distribution Functions}

The $x$-dependence of the structure functions carries valuable information
about the dynamics of quarks and gluons which is not immediately
available from the moments. Furthermore, because of limited
experimental data, moments are
sometimes hard to compare with experiment. This is, in particular, the
case for the higher moments. Theoretically, the structure functions can
be obtained from the moments by an inverse Mellin transform.

A first attempt of constructing nucleon structure functions from a few
lower moments was reported in \cite{mw} for the unpolarized case. \hfill In
this talk we shall consider a\\ 

\clearpage
\begin{figure}[h]
\begin{centering}
\epsfig{figure=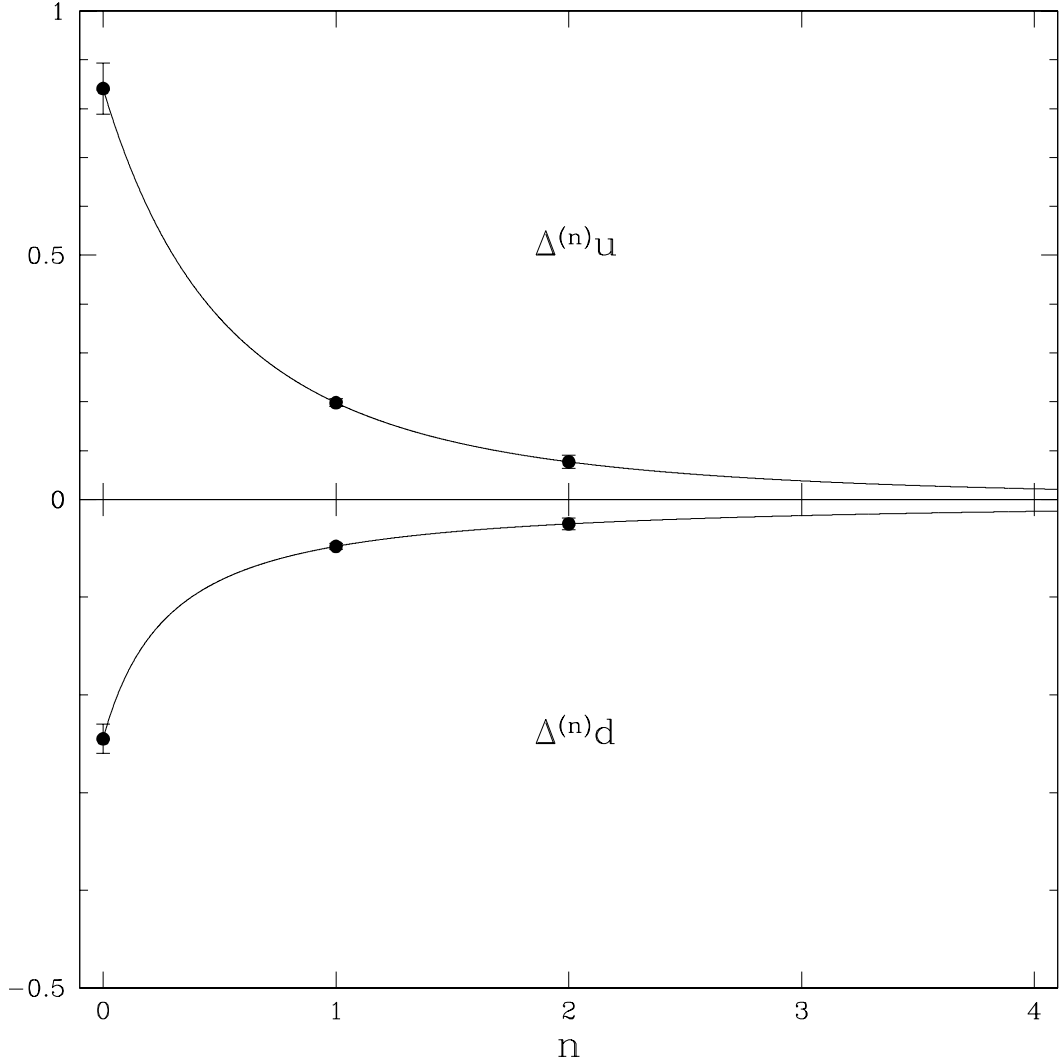,height=15.5cm,width=15.5cm}
\caption{The lattice results for the first three moments of $\Delta u(x)$
and $\Delta d(x)$ at $\beta = 6.0$ and $\mu^2 = 4 \; \mbox{GeV}^2$. 
The curves are fits to the lattice data of the form
(\protect\ref{fit}). The parameters of the fit are $\alpha = 0.04(6)$
and $\beta = 2.21(9)$.}
\label{moms}
\end{centering}
\end{figure}

\clearpage
\begin{figure}[h]
\begin{centering}
\epsfig{figure=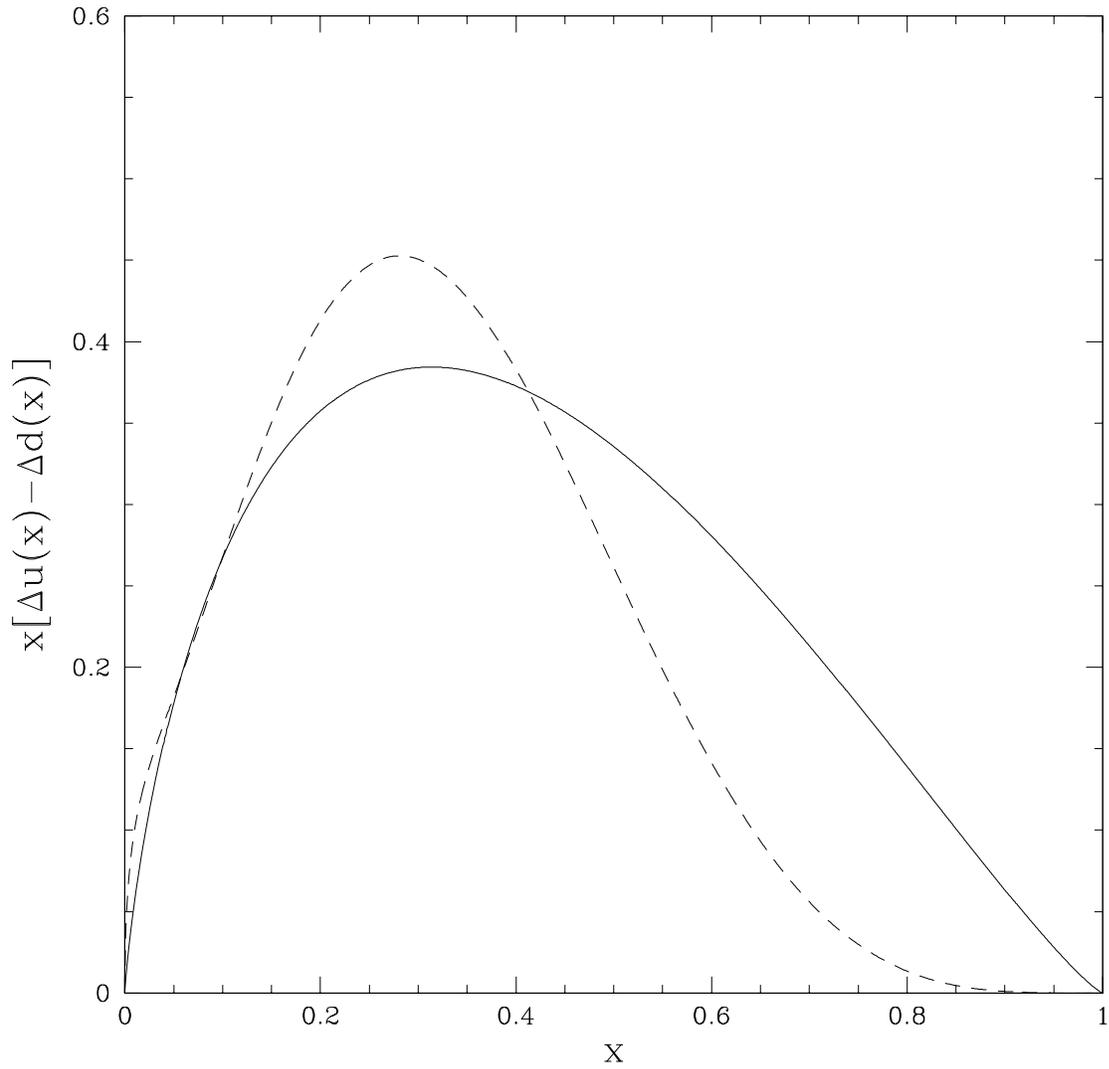,height=15.5cm,width=15.5cm}
\caption{The distribution $x[\Delta u(x) - \Delta d(x)]$, together with
the phenomenological valence quark distribution. The solid
line is the result of the fit, the dashed line is taken from 
ref.~\protect\cite{gs}.}
\label{stru}
\end{centering}
\end{figure}

\clearpage

\noindent
different method and apply it to $g_1$.

We restrict ourselves to the coarser lattice at $\beta = 6.0$. We have
computed the three lowest moments. Calculation of a fourth moment is
possible. The moments can be well described by the formula
\begin{equation}
\Delta^{(n)} q = c \: \Gamma(\beta) (n + 1 + \alpha)^{-\beta}.
\label{fit}
\end{equation}
This formula seems also to describe the moments of the unpolarized
structure functions well. 
A fit is shown in Fig.~\ref{moms}. An inverse Mellin transform
of (\ref{fit}) gives
\begin{equation}
\Delta q(x) = c \: x^\alpha (-\ln x)^{\beta - 1}.
\end{equation}
In Fig.~\ref{stru} we compare the result for $\Delta u(x) - \Delta d(x)$
with the phenomenological distributions. The outcome is
encouraging. We have hope that with one more moment and precise
lattice data we are able to derive phenomenologically useful quark
distribution functions.   

It must be said that one can only combine even and odd moments to make
a single structure function if $u$ and $d$ sea quark contributions are
assumed to be equal, as is commonly done~\cite{gs}. However, this may
not be a good approximation~\cite{ross}.

\section{Conclusions}

Lattice calculations of nucleon structure functions have improved in
many respects. The calculations are now done with improved actions and
using improved operators, so as to reduce cut-off
effects. Furthermore, the renormalization constants of the lattice
operators, which are another source of errors, are gradually being computed  
non-perturbatively~\cite{alpha,paul}. On top of that, we have seen
that it is important to do the calculation at several values of the
coupling and do an extrapolation to $a = 0$. By adding one or two more data
points, and with increased statistics, we will soon be ready to report
reliable continuum results, at least in
the quenched approximation.

Where we can compare the lattice results with experiment or the
phenomenological analysis, we find good agreement.
Our efforts over the last year have, in particular, paid off for
$g_A$, the axial vector
coupling of the nucleon.    
Two years ago this quantity was considered a problem for quenched lattice 
QCD~\cite{okawa}.

\end{document}